# Structured air lasing of $N_2^+$

Jingsong Gao[1], Xiang Zhang[2], Yang Wang[1], Yiqi Fang[1], Qi Lu[2], Zheng Li[1], Yi Liu[2,*], Chengyin Wu[1], Qihuang Gong[1], Yunquan Liu[1,†], Hongbing Jiang[1,‡]

[1]*State Key Laboratory for Mesoscopic Physics, School of Physics, Peking University, Beijing 100871, China*

[2] *Shanghai Key Lab of Modern Optical System, University of Shanghai for Science and Technology, Shanghai 200093, China*

*yi.liu@usst.edu.cn

†yunquan.liu@pku.edu.cn

‡hbjiang@pku.edu.cn

## Abstract

Structured light has attracted great interest in scientific and technical fields. Here, we demonstrate the first generation of structured air lasing in $N_2^+$ driven by 800 nm femtosecond laser pulses. By focusing a vortex pump beam at 800 nm in $N_2$ gas, we generate a vortex superfluorescent radiation of $N_2^+$ at 391 nm, which carries the same photon orbital angular momentum as the pump beam. With the injection of a Gaussian seed beam at 391 nm, the coherent radiation is amplified, but the vorticity is unchanged. A new physical mechanism is revealed in the vortex $N_2^+$ superfluorescent radiation: the vortex pump beam transfers the spatial spiral phase into the $N_2^+$ gain medium, and the Gaussian seed beam picks up the spatial spiral phase and is then amplified into a vortex beam. Moreover, when we employ a pump beam with a cylindrical vector mode, the Gaussian seed beam is correspondingly amplified into a cylindrical vector beam. Surprisingly, the spatial polarization state of the amplified radiation is identical to that of the vector pump beam regardless of whether the Gaussian seed beam is linearly, elliptically, or circularly polarized. Solving three-dimensional coupled wave equations, we show how a Gaussian beam becomes a cylindrical vector beam in a cylindrically symmetric gain medium. This study provides a novel approach to generating structured light via $N_2^+$ air lasing.

## Introduction

With the development of emerging laser sculpting techniques, the amplitude, phase and polarization of the fundamental Gaussian laser mode can be completely controlled, giving rise to many interesting spatiotemporal modes[1], such as optical vortices (OVs)[2] and vector beams (VBs)[3]. An OV is a light field with a spatially spiral phase structure and intrinsic photon orbital angular momentum (OAM), and it is typically described by a Laguerre-Gaussian (LG) mode[4]. Because the OAM of a photon is multiple-valued, it provides a highly dimensional degree of freedom for applications in many fields, such as super-resolution microscopy[5], optical tweezers[6,7], spinning object detection[8], and high-capacity optical communication via OAM multiplexing[9,10]. In contrast, a VB is a light field with a spatially varying state of polarization. One paradigm of VBs is cylindrical vector beams (CVBs), which exhibits cylindrically symmetric polarization[3]. Owing to their unique dynamic properties, CVBs have attracted broad interest in various scientific communities, ranging from optical trapping[11] and high-resolution imaging[12,13] to communications[14,15] and quantum memory[16]. In particular, cylindrical vector vortex beams, that is, CVBs carrying OAM, have recently been used to study the spin-orbit interaction of light[17,18]. The generation of OVs and CVBs conventionally relies on linear optical processes and photonic devices, such as spiral phase plates (SPPs)[19], spatially variant half-wave plates (SWPs, half-wave q-plates)[20], and spatial light modulators[21,22]. Recently, nonlinear intense light-matter interactions have propelled the generation of structured light fields with an extensive wavelength range. For example, one can employ high harmonic generation to create time-varying OAM[23] and simultaneously control the spin-orbit state of extreme ultraviolet photons[24]. Indeed, it is worthwhile to seek new generation routines of structured light fields using light-matter interactions, which would provide a robust and promising toolbox for complete control over structured light fields.

Air lasing is a cavity-free lasing action that is generated in air owing to the plasma filamentation process of high-power femtosecond laser pulses[25,26]. In particular, the coherent emission of $N_2^+$ at 391 nm pumped by femtosecond laser pulses at 800 nm has gained increasing attention due to its rich physical mechanism[27–36]. It is now generally accepted that the coherent emission of $N_2^+$ at 391 nm is essentially superfluorescence[34,35], which is a distinctive quantum optics phenomenon[37-39] and is collectively emitted from a macroscopic dipole[40]. If an external seed light with a spectrum covering 391 nm is injected into the plasma, the coherent emission of $N_2^+$ at 391 nm will be amplified by several orders of magnitude[35]. Many studies on enhancing the intensity of $N_2^+$ air lasing are ongoing[41,42]. However, laser beams with space-variant phases or polarizations, that is, OVs or VBs, have not yet been applied to $N_2^+$ air lasing.

Here, we report a novel physical phenomenon in structured $N_2^+$ air lasing. Spatially structured light is first employed to drive $N_2^+$ air lasing. By focusing a *p*-polarized vortex pump beam at 800 nm on $N_2$ gas, we generate a vortex signal of $N_2^+$ air lasing at 391 nm with the same OAM as the pump beam. With the injection of a *p*-polarized Gaussian seed beam with an optimal delay, the coherent signal is amplified by nearly three orders of magnitude. The amplified signal is also an OV with the same OAM. This indicates that the vortex pump beam transfers the spatial spiral phase to the $N_2^+$ gain medium, and the Gaussian seed beam picks up the spatial spiral phase and is then amplified into an OV. In the presence of the seed beam, we change the vortex pump beam to a CVB pump beam. Interestingly, regardless of whether the polarization state of the seed beam is linear, elliptical, or circular, the amplified signal will always maintain the same CVB mode as the pump beam. We numerically solve three-dimensional coupled wave equations and show how a Gaussian beam eventually evolves into a CVB mode in a cylindrically symmetric gain medium.

## Results

The schematic of the experiment is depicted in Fig. 1 (see Methods for more details of the experimental setup). We utilize a SPP to transform *p*-polarized intense femtosecond laser pulses at 800 nm into an OV with a topological charge of $\ell = 1$ (photon's OAM is $\ell\hbar$). As shown in Fig. 1a, the OV is used as the pump beam to ionize $N_2$ molecules. The ionization occurs in a laser plasma filament when the pump beam is focused by a lens with a focal length of 30 cm. Nitrogen molecules are mainly ionized into three electronic states of $N_2^+$. As shown in the left inset of Fig. 1a, the three states constitute a V-type three-level system consisting of $X^2\Sigma_g^+$, $A^2\Pi_u$ and $B^2\Sigma_u^+$ states.

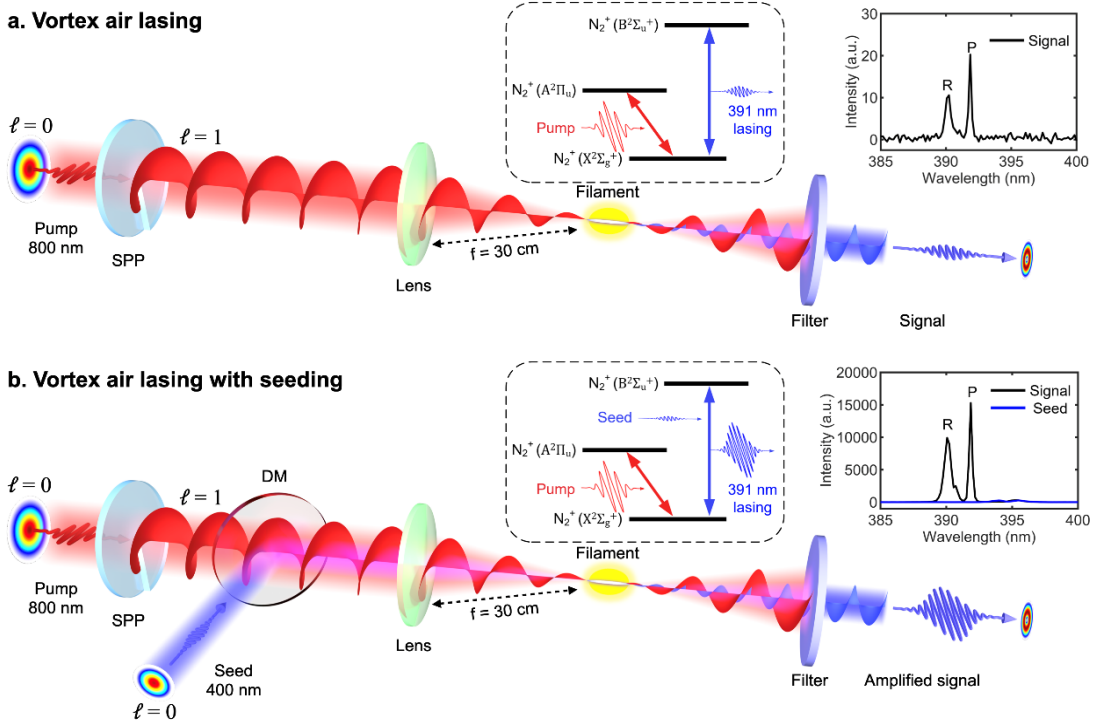

**Fig. 1 Schematic of the experiment.** Layout diagraming the generation processes of vortex air lasing of $N_2^+$ without seeding (**a**) and with seeding (**b**). **a**, an intense vortex beam at 800 nm is used to pump $N_2$ molecules alone. **b**, an intense vortex beam at 800 nm is used to pump $N_2$ molecules and a weak Gaussian beam with a spectrum covering 391 nm is injected to stimulate excited $N_2^+$ ions. Left insets surrounded by dash lines: schematics of the three-level system of $N_2^+$ air lasing at 391 nm without seeding (**a**) and with seeding (**b**). Right insets: spectra of (**a**) the signal of $N_2^+$ air lasing; (**b**) the amplified signal of $N_2^+$ air lasing and the seed light. The filter is a bandpass filter with a center wavelength of 390 nm and full width half maximum of 10 nm.

The spectrum of the detected signal is plotted in the right inset of Fig. 1a, represented by the black line. The signal is identified as coherent radiation of $N_2^+$ air lasing at 391 nm, which is emitted by the transition of $B^2\Sigma_u^+(v''=0) \to X^2\Sigma_g^+(v=0)$ and includes the P branch (rotation transition: $J \to J+1$) and the R branch (rotation transition: $J \to J-1$). The beam profile of the signal is a doughnut-type and is recorded by a charge-coupled device (CCD), as shown in Fig. 2b. The signal has the same *p*-polarized electric field as the pump beam at each spatial point. The singularity at the beam center originates from the helical phase structure. To measure the topological charge of the signal, we employ momentum space mapping of a cylindrical lens[41–43] which is schematically depicted in Fig. 2a. The pattern of twisted light at the focus of a cylindrical lens is OAM-dependent. The simulated images at the focal plane of a cylindrical lens are illustrated in Fig. 2f, in which the OVs become deformed patterns with $\ell + 1$ skew bright stripes. Two adjacent bright stripes constitute a deformed vortex core, and the number of the cores is equal to the topological charge. The signal beam is focused experimentally using a cylindrical lens with a focal length of 1 m. As shown in Fig. 2c, the corresponding focal image is a pattern with two bright stripes. Compared to the simulation results, it is evident that the signal beam is an OV with a topological charge of $\ell = 1$. This implies that the spatial phase structure of the pump beam has been entirely imprinted on the coherent signal of $N_2^+$ at 391 nm.

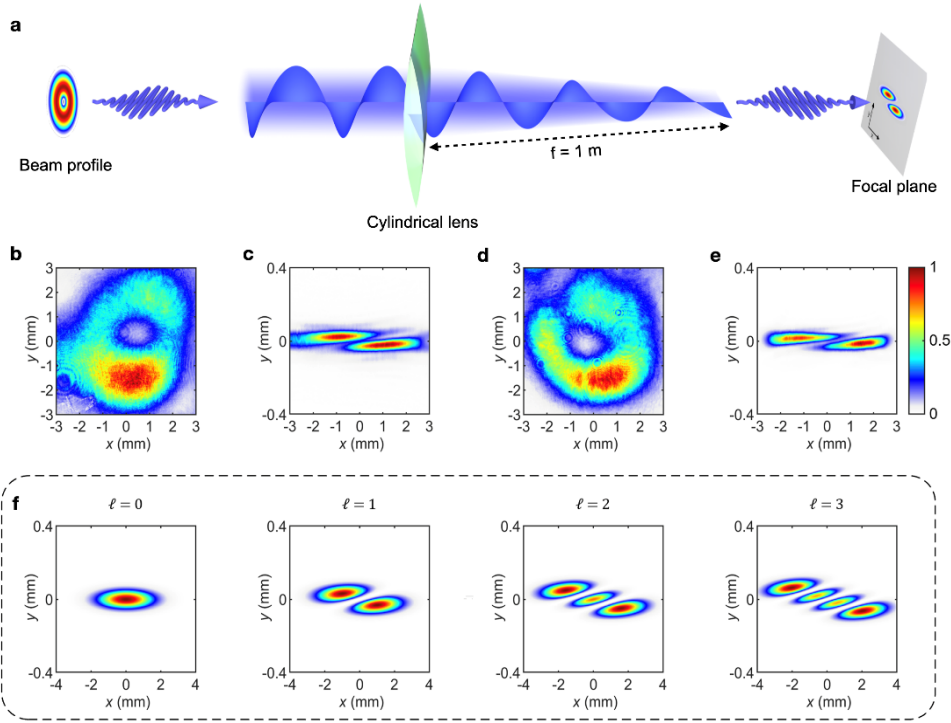

**Fig. 2 Intensity distributions and focal images of $N_2^+$ air lasing at 391 nm pumped by OVs. a**, Schematic of the momentum space mapping of a cylindrical lens. **b-c**, The measured intensity distribution and focal image of $N_2^+$ air lasing at 391 nm without seeding. **d-e**, The measured intensity distribution and focal image of $N_2^+$ air lasing at 391 nm with seeding. **f**, Simulated focal images of vortex beams with different topological charges, that is, 0, 1, 2 and 3, respectively. All beams are focused in the y direction by the cylindrical lens. The focal length of the cylindrical lens is 1 m in both experiments and calculations.

With the injection of a *p*-polarized Gaussian seed beam at an optimal delay (the seed beam has a central wavelength of 400 nm and a spectrum covering 391 nm), the intensity of the signal is amplified, as shown in Fig. 1b. The amplified signal is nearly three orders of magnitude stronger than the unseeded signal. The spectrum of the amplified signal is represented by the black line in the right inset of Fig. 1b. As shown in Fig. 2d, the intensity distribution of the amplified signal also has a doughnut shape. Similarly, the amplified signal beam is focused by the cylindrical lens, and the focal image is illustrated in Fig. 2e. This reveals that the amplified signal also has the same topological charge as that of the pump beam. However, in the theory of superfluorescence, it is commonly regarded that if an external seed is injected, it will dominate the establishment of macroscopic polarization. Theoretically, the spatial phase distribution of the amplified signal should be independent of that of the vortex pump beam. Interestingly, our results indicates that the spatial phase of the pump beam is simultaneously transferred to the amplified signal via the $N_2^+$ gain medium, which points to a new physical mechanism in $N_2^+$ air lasing in addition to the current superfluorescence theory.

Inspired by the results of OVs with seeding, we replace the SPP with an SWP to change the pump beam into a CVB. The azimuthal angle of the SWP is rotated to let the pump beam become radially polarized. In the same way, the *p*-polarized Gaussian seed beam is injected into the plasma filament with an optimal delay, and then an amplified signal of $N_2^+$ air lasing at 391 nm is observed. As shown in the plot titled $|E|^2$ in Fig. 3a, the intensity distribution of the amplified signal is also a doughnut-shaped pattern with clear boundaries. In contrast to OVs, spatially variant polarization states contribute to the singularity here. A polarizer is placed in front of the CCD to observe the spatial polarization distribution. The plots of $|E_x|^2$, $|E_x+E_y|^2/2$, and $|E_y|^2$ in Fig. 3a show the intensity distributions of the amplified air lasing after passing through the polarizer with different polarization axis angles, that is, 0°, 45°, and 90°, respectively. As can be seen, if filtered by a polarizer, the intensity distribution of the amplified signal beam will appear as two lobes that rotate with the polarization axis angle of the polarizer. The *x* and *y* components, $|E_x|^2$ and $|E_y|^2$, possess the features of $HG_{01}$ and $HG_{10}$ modes, respectively (HG modes, Hermit-Gaussian modes[44]). It is well known that a radially polarized beam can be expressed as a coherent superposition of *p*-polarized $HG_{01}$ and *s*-polarized $HG_{10}$ modes[3]. The results show that the amplified signal exhibits the characteristics of a radially polarized mode, which is identical to that of the pump beam. We then change the seed beam into elliptically and circularly polarized states in turn (the ellipticities are 0.73 and 0.92, respectively), and the corresponding results have the same characteristic of the radially polarized mode, as shown in Fig. 3b and Fig. 3c. It is obvious that the polarization state of the amplified signal is completely dictated by the vector pump beam and is independent of the Gaussian seed beam.

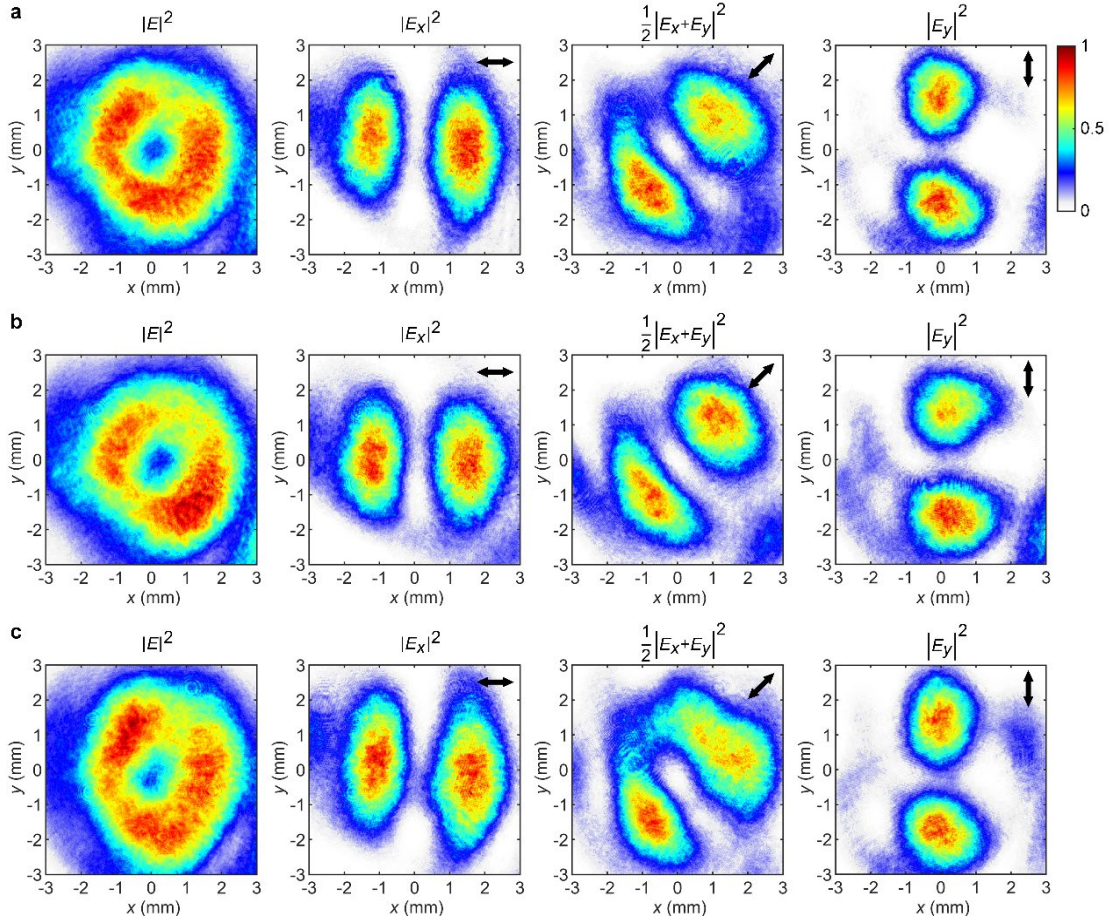

**Fig. 3 Intensity distributions of $N_2^+$ air lasing at 391 nm pumped by a radially polarized CVB and stimulated by external Gaussian seed beams. a-c,** The polarization states of seed beams are linearly, elliptically, and circularly polarized, respectively. The panels titled $|E|^2$ show the intensity distributions of signals directly recorded by the CCD. The panels titled $|E_x|^2$, $\frac{1}{2}|E_x+E_y|^2$, and $|E_y|^2$ show the intensity distributions filtered by a polarizer with different angles, that is, 0°, 45° and 90°, respectively. The intensity distributions of each row are normalized by the same standard as the first panel. The black arrows indicate the polarization axis direction of the polarizer.

For further confirmation, we rotate the SWP to allow the pump beam to become azimuthally polarized. Similar to the results shown in Fig. 3, regardless of whether the seed beam is linearly, elliptically, or circularly polarized, the corresponding amplified signals always process the feature of the azimuthally polarized mode, as shown in Fig. 4. Moreover, if the pump beam is in any CVB mode that is between the radially and azimuthally polarized modes, the amplified signal will also possess the same CVB mode as the pump beam. Therefore, a CVB mode tuning of $N_2^+$ air lasing can be simply implemented.

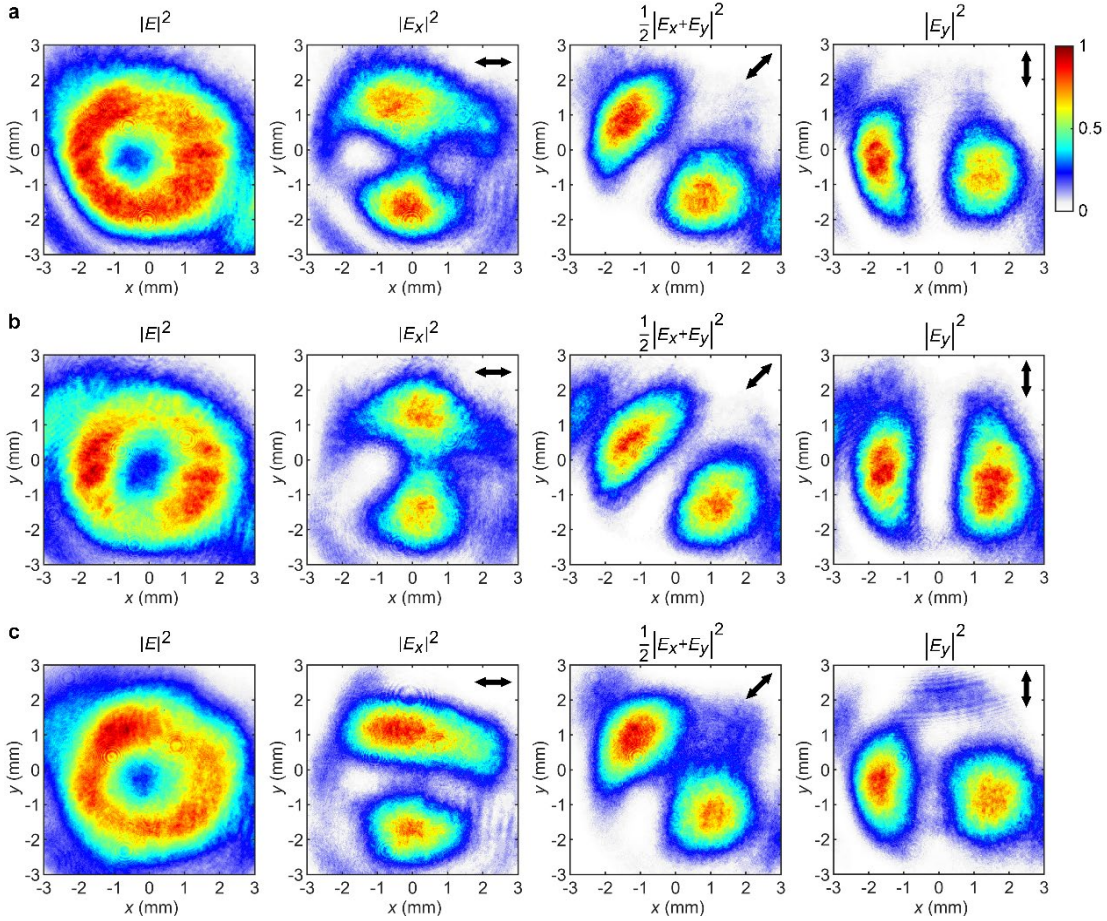

**Fig. 4 Intensity distributions of $N_2^+$ air lasing at 391 nm pumped by an azimuthally polarized CVB and stimulated by external Gaussian seed beams. a-c,** The polarization states of seed beams are linearly, elliptically, and circularly polarized, respectively. The panels titled $|E|^2$ show the intensity distributions of signals directly recorded by the CCD. The panels titled $|E_x|^2$, $\frac{1}{2}|E_x+E_y|^2$ and $|E_y|^2$ show the intensity distributions filtered by a polarizer with different angles, that is, 0°, 45° and 90°, respectively. The intensity distributions of each row are normalized by the same standard as the first panel. The black arrows indicate the polarization axis direction of the polarizer.

## Discussion

To interpret how a Gaussian beam can be amplified into a CVB, we build a theoretical model of a cylindrically symmetric gain medium. In previous experiments, the polarization states of the pump beams and seed beams are usually linearly polarized and synergistically determine the polarization direction of the amplified signal[45]. The amplified signal is linearly polarized, and the polarization direction is between the pump and seed beams. In addition, the intensity of the amplified signal is dependent on the angle between the polarization directions of the pump and seed beams. When the angle is 0°, the amplification ratio is maximum[46]. It decreases as the angle increases, until it reaches a minimum when the angle is 90°. This implies that the maximum and minimum gain directions are parallel and perpendicular to the polarization direction of the pump beam, respectively. We believe that this is caused by the average alignment of $N_2^+$ induced by the pump beams. First, the pump beams align the nitrogen molecules along the laser polarization direction before and after the ionization. In addition, the optimal ionization direction of the highest occupied molecular orbital (HOMO) and HOMO-2 of $N_2$ is parallel to the molecular axis[47]. If $N_2$ molecules eject an outer valence electron in HOMO or HOMO-2, they will become $N_2^+$ ions in the state of $X^2\Sigma_g^+$ or $B^2\Sigma_u^+$[48]. These two factors result in most of the $N_2^+$ ions in the states $X^2\Sigma_g^+$ and $B^2\Sigma_u^+$ being aligned along the polarization direction of the pump beams[49]. Because the transition dipole moment between $X^2\Sigma_g^+$ and $B^2\Sigma_u^+$ is also parallel to the molecular axis[29], the electric field component of the seed beam parallel to the polarization direction of the pump beam has the maximum amplification, and the component perpendicular to the polarization direction of the pump beam has the minimum amplification.

Consequently, if the pump beams are radially (azimuthally) polarized, the maximum gain direction of the $N_2^+$ medium will present a radial (azimuthal) distribution in the filament; hence, the minimum gain direction will be azimuthal (radial). Based on this, we build a theoretical model of the medium with a cylindrically symmetric gain, which is described by two three-dimensional coupled wave equations (see Methods for details of the model). In this model, we focus on how the signal spatially evolves from a Gaussian mode into a CVB mode in an anisotropic gain medium. Thus, the time dependence of the population distribution in the ground and excited states of $N_2^+$ is neglected, i.e., the amplification term is deemed as a constant in the calculations. The effect of the photoinduced refractive index change can also be ignored, because it has no impact on the cylindrical symmetry of the results.

The numerical results of the theoretical model are shown in Fig. 5. This shows an evolution process in which a *p*-polarized Gaussian seed beam gradually becomes a radially polarized beam as it propagates in a radial gain medium. When the propagation distance is 0 mm, the *y* component of the intensity is null, and the intensity distribution, $|E|^2$, is identical to its *x* component, $|E_x|^2$. As the propagation distance increases, both the intensity of the light field and the ratio of the *y* component to the *x* component increase. At the exit of the gain medium, the seed beam has been amplified thousands of times and eventually becomes radially polarized, as shown in Fig. 5c. Similarly, if the gain distribution of the medium is azimuthal, the *p*-polarized seed beam will be amplified into an azimuthally polarized beam. The model is applicable to any polarization state of the seed beam. We also simulate the injection of a circularly polarized seed beam, which is also amplified into a CVB.

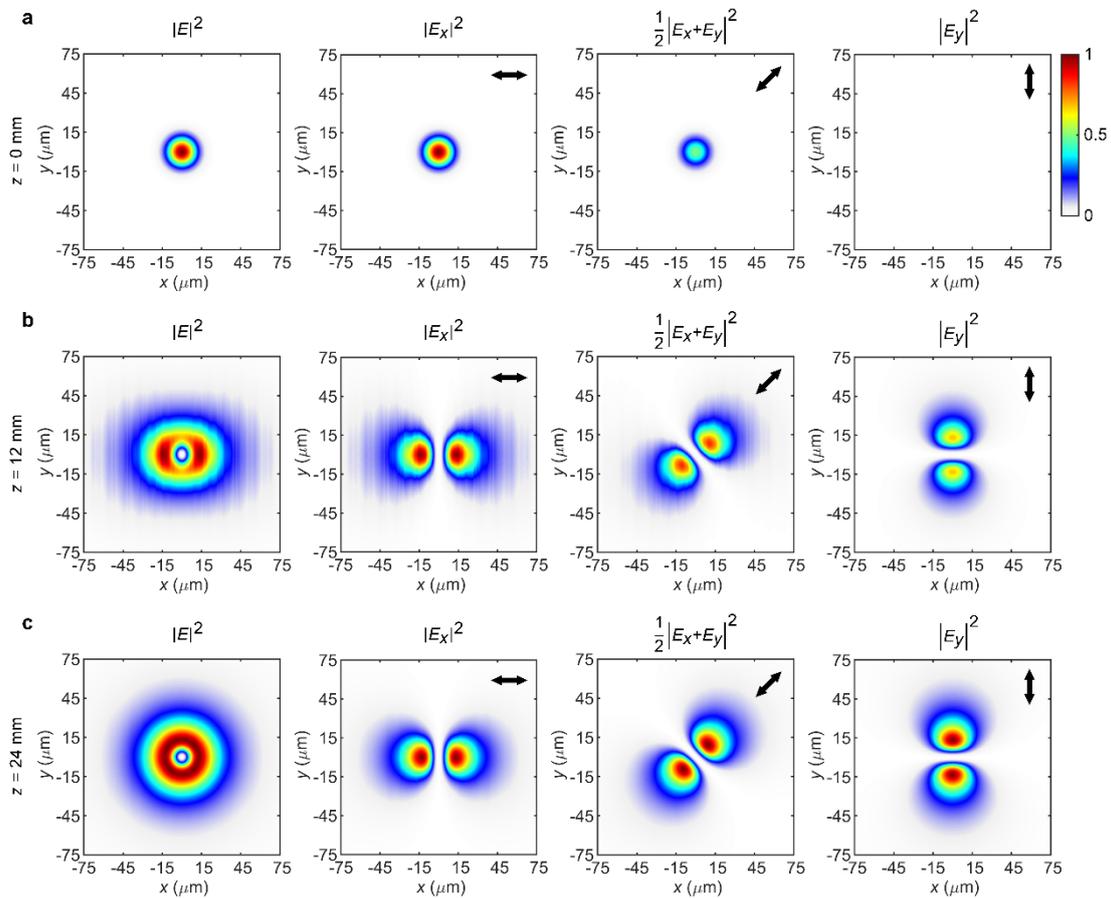

**Fig. 5 Simulated distributions of light intensity in a medium with radial gain stimulated by a *p*-polarized Gaussian seed beam. a-c**, Simulated intensity distributions at different distances in the gain medium, that is, 0 mm, 12 mm and 24 mm, respectively. The amplification ratios are about 1 (**a**), 154 (**b**) and 6647 (**c**) times, respectively. The intensity distributions of each row are normalized by the same standard as the first panel. The black arrows indicate the directions of polarization components.

Generally, the cylindrically symmetric gain medium supports a feasible way in which one component of the electric field can be amplified to that is orthogonal to it, namely, the $E_y$ can be amplified from the $E_x$. However, the anisotropic gain medium cannot provide an abrupt transverse phase difference of π between the centrosymmetric points for the formation of a CVB when it is seeded by a Gaussian beam. Interestingly, the transfer of the spatial phase from the pump beam to the amplified signal can compensate for the phase difference. In the calculations, we treat the phase-transfer problem by imposing the spatial phase of the pump beam on the seed beam at the entrance of the gain medium. Thus, a weak Gaussian seeding beam will evolves into a strong CVB during the propagation of a medium with a cylindrically symmetric gain.

In air lasing using a vortex pump beam without seeding, the coherent signal of $N_2^+$ air lasing at 391 nm inherits the spiral phase structure of the vortex pump beam. This is consistent with the previous understanding that $N_2^+$ air lasing at 391 nm is self-seeded by the white light originating from the supercontinuum of the pump beam. The supercontinuum in the filament extends the spectrum of the pump beam from the infrared region to the ultraviolet region as a result of self-phase modulation[50], which naturally brings the helical phase structure of the pump light to the white light.

When seeding with a Gaussian beam, the coherent signal is amplified and still has the same topological charge as the pump beam. One may expect that, in the current theory of superfluorescence, the spatial phase information of the pump beam will be lost in the amplified signal under the V-type three-level system because it is generally considered that the macroscopic polarization of the $N_2^+$ medium is predominantly induced by the external seed beam. Theoretically, the spatial phase distribution of the amplified signal should be the same as that of the Gaussian seed beam. Interestingly, the spatial spiral phase of the pump beam is eventually transferred to the amplified signal of $N_2^+$ air lasing at 391 nm, namely, the OAM of the pump beam is transferred to the amplified signal via $N_2^+$ medium in the laser filamentation. This manifests that there is a new physical mechanism in $N_2^+$ air lasing in addition to the current superfluorescence theory, which needs to be further studied.

If pumping with a CVB in the presence of seeding, the Gaussian seed beam is correspondingly amplified into a CVB. The polarization state of the amplified signal is completely determined by the vector pump beam. Regardless of whether the seed beam is linearly, elliptically, or circularly polarized, the amplified signal will always be a CVB with an identical mode as the pump beam. The vector results with seeding are different from previous experimental results, where the pump and seed beams are both linearly polarized[45]. In the previous cases, the pump and seed beams synergistically control the polarization direction of the amplified signal. Consequently, the amplified signal is linearly polarized, and the polarization direction is between the pump and seed beams. These two different phenomena both originate from the laser alignment of $N_2^+$ induced by the pump beam, i.e., the laser alignment determine the spatial feature of maximum gain direction of $N_2^+$ medium. While a *p*-polarized pump beam produces a $N_2^+$ medium with a horizontal maximum gain direction, the maximum gain direction of the $N_2^+$ medium induced by a CVB exhibits a cylindrical distribution. The vector results suggests that laser-induced alignment could potentially be employed in commercial molecular lasers to generate high-power CVBs or other kinds of VBs.

In conclusion, we report the first implementation of structured air lasing. We have experimentally generated two typical types of structured light by $N_2^+$ air lasing, that is, OVs and CVBs. We find that the OV at 800 nm transfers the spatial spiral phase to the $N_2^+$ gain medium, and the seed beam picks up the spatial spiral phase and is then amplified into a vortex beam at 391 nm. The phase transfer from the pump beam to the seed beam suggests a novel mechanism of $N_2^+$ air lasing that deserves further study. The CVB at 800 nm aligns the nitrogen ions along its polarization direction, producing a cylindrically symmetric gain medium of $N_2^+$. This anisotropic gain medium allows for the amplification of a seed beam with any spatially homogeneous polarization states into a CVB. From a technical perspective, this work provides a promising approach to generating structured light, in which a Gaussian beam can be directly amplified into structured light by cavity-free lasing action instead of a complicated optical system.

## Methods

**Experimental details.** The *p*-polarized Gaussian laser pulses at a central wavelength of 800 nm are launched from a regenerate amplification system of a Ti:Sapphire laser, which delivers 35 fs pulses at a repetition frequency of 1 kHz. The laser pulses are then split into two beam paths. The first one is used as a pump beam with a single-pulse energy of 1.8 mJ. First, an SPP is placed in the beam path to transform the pump beam into a vortex beam (the topological charge is $\ell = 1$). Then, we turn the pump beam into a CVB by replacing the spiral phase plate with a SWP. By rotating the azimuthal angle of the SWP, the CVB mode can be continuously tuned from radial polarization to azimuthal polarization. The second path passes through a half-wave plate and a 0.1-mm-thick beta-barium borate crystal (BBO) to produce a *p*-polarized second harmonic with a central wavelength of 400 nm and a spectrum covering 391 nm. The second harmonic is used as a weak seed beam and a quarter-wave plate is utilized to change its polarization state (linear, elliptical or circular polarization). The two beams are focused collinearly into a gas chamber using a lens with a focal length of 30 cm and a single filament is produced. The chamber is filled with pure $N_2$ gas at 30 mbar. The pump beam is filtered out by two bandpass optical filters before signal detection. The lasing emission of $N_2^+$ at 391 nm is collected by a fiber head connected to a grating spectrometer or recorded by a CCD.

**Theoretical model.** The three-dimensional coupled wave equations of the medium with radial gain (under the condition of a slowly varying approximation) are expressed as

$$\frac{\partial^2 E_x}{\partial x^2} + \frac{\partial^2 E_x}{\partial y^2} + 2ik\frac{\partial E_x}{\partial z} = -\mu\omega^2 P_x, \qquad (1)$$

and

$$\frac{\partial^2 E_y}{\partial x^2} + \frac{\partial^2 E_y}{\partial y^2} + 2ik\frac{\partial E_y}{\partial z} = -\mu\omega^2 P_y, \qquad (2)$$

where $E_x$ and $E_y$ are the *x* and *y* components of the slowly varying amplitude of the electric field, respectively, $k$ is the wave number of the electric field, $\omega$ is the angular frequency of the electric field, $\mu$ is the permeability, $P_x$ and $P_y$ are the *x* and *y* components of the slowly varying part of the medium macroscopic polarization, respectively. The macroscopic polarization is the functions of both $E_x$ and $E_y$, namely,

$$P_x = C\langle\cos^2\theta\rangle(E_x\cos^2\varphi + E_y\sin\varphi\cos\varphi)$$
$$+ C\langle\sin^2\theta\rangle(E_x\sin^2\varphi - E_y\sin\varphi\cos\varphi), \qquad (3)$$
$$P_y = C\langle\cos^2\theta\rangle(E_x\sin\varphi\cos\varphi + E_y\sin^2\varphi)$$
$$- C\langle\sin^2\theta\rangle(E_x\sin\varphi\cos\varphi - E_y\cos^2\varphi). \qquad (4)$$

The parameter $C$ is given by

$$C = -\frac{i}{\hbar\gamma}d_{BX}^2 N(\rho_{22} - \rho_{11}), \qquad (5)$$

where $\hbar$ is the reduced Planck constant, $\gamma$ is the decay rate, $d_{BX}$ is the transition dipole moment between $X^2\Sigma_g^+$ and $B^2\Sigma_u^+$ states, $N$ represents the average number of particles per unit volume, $\rho_{11}$ and $\rho_{22}$ are the probabilities of being in the upper and lower states, respectively, $N(\rho_{22} - \rho_{11})$ represents the difference in the population of the upper and lower states per unit volume that is determined by the intensity of pump light and characterizes the amplification, $\varphi$ is the azimuth of the *xy* plane, and $\langle\cos^2\theta\rangle$ and $\langle\sin^2\theta\rangle$ are the average alignment degrees of $N_2^+$ that are parallel and perpendicular to the polarization direction of the pump beam, respectively (the details of the derivation will be provided with the official publication in the form of supplementary information). The equations are solved using the split-step Fourier method. In the calculations, the time dependence of the population distribution in the ground and the excited states is neglected, namely, $\rho_{22}$ and $\rho_{11}$ are regarded as constants. The waists of the seed beam and the gain medium (plasma filament) are both set to 15 μm. The cross section of the gain medium is set as the LG mode with $\ell = 1$ (because the intensity distribution of CVBs has the same pattern as that of the OV with $\ell = 1$). The ratio of $\langle\cos^2\theta\rangle$ to $\langle\sin^2\theta\rangle$ is set to 2. the length of the gain medium is set to 24 mm. The parameter $C$ is set to make the seed light amplified around 8000 times. If the gain distribution is azimuthal, then $\langle\cos^2\theta\rangle$ and $\langle\sin^2\theta\rangle$ will exchange with each other in the equations.

## Acknowledgments

This work is supported by the National Key R&D Program of China (Grant No. 2018YFA0306302), and the National Natural Science Foundation of China (Grants Nos. 41527807, 12034013)